\documentclass[iop,onecolumn]{emulateapj}

\accepted{\today}
\shorttitle{Dynamic evolution and fractal tearing in current sheets}
\shortauthors{Singh et al.}
\usepackage{natbib}
\usepackage{hyperref}
\usepackage{xcolor}
\usepackage{amsmath}
\usepackage{rotating}
\usepackage{array}
\usepackage{bibentry}
\usepackage{amsmath}
\usepackage{enumitem}
\usepackage{booktabs}
\usepackage{multirow}
\usepackage{gensymb}
\usepackage{enumerate}
\usepackage{epstopdf}

\defcitealias{PV14}{PV14}

\begin{document}

\title{Dynamic evolution of current sheets, ideal tearing, plasmoid formation and generalized fractal reconnection scaling relations}

\author{K.A.P. Singh}
\affiliation{Department of Physics, Institute of Science, BHU, Varanasi 221005, India}
\affiliation{Astronomical Observatory, Graduate School of Science, Kyoto University, Yamashina, Kyoto 607-8471, Japan}
\author{Fulvia Pucci}
\affiliation{National Institutes for Natural Sciences, Toki, Japan}
\affiliation{Princeton Plasma Physics Laboratory, Princeton University, Princeton, New Jersey 08543-0451, USA}

\author{Anna Tenerani}
\affiliation{Department of Physics, The University of Texas at Austin, TX 78712, USA}

\author{Kazunari Shibata}
\affiliation{Astronomical Observatory, Graduate School of Science, Kyoto University , Yamashina, Kyoto 607-8471, Japan}

\author{Andrew Hillier}
\affiliation{Department of Mathematics, CEMPS, University of Exeter, Exeter EX4 4QF, UK}

\author{Marco Velli}
\affiliation{Department of Earth, Planetary, and Space Sciences, UCLA, Los Angeles, CA 90095, USA}

\begin{abstract}

Magnetic reconnection may be the fundamental process allowing energy stored in magnetic fields to be released abruptly, solar flares and coronal mass ejection (CME) being archetypal natural plasma examples.  Magnetic reconnection is much too slow a process to be efficient on the large scales, but accelerates once small enough scales are formed in the system. For this reason, the fractal reconnection scenario was introduced (Shibata and Tanuma 2001) to explain explosive events in the solar atmosphere: it was based on the recursive triggering and collapse via tearing instability of a current sheet originally thinned during the rise of a filament in the solar corona.  Here we compare the different fractal reconnection scenarios that have been proposed, and derive generalized scaling relations for the recursive triggering of  fast,  `ideal' - i.e. Lundquist number independent - tearing in collapsing current sheet configurations with arbitrary current profile shapes. An important result is that the Sweet-Parker scaling with Lundquist number, if interpreted as the aspect ratio of the singular layer in an ideally unstable sheet, is universal and does not depend on the details of the current profile in the sheet. Such a scaling however must not be interpreted in terms of stationary reconnection, rather it defines a step in the accelerating sequence of events of the ideal tearing mediated fractal cascade. We calculate scalings for the expected number of plasmoids for such generic profiles and realistic Lundquist numbers.

\end{abstract}

\keywords{magnetic reconnection--- magnetohydrodynamics (MHD)---plasmas}

\section{Introduction} \label{sec:intro}

Magnetic reconnection is a dynamical mechanism pervasive in the high temperature, low resistivity plasmas common in astrophysical settings as well as in the laboratory in fusion research. It is considered to be one of the most fundamental processes permitting mass, momentum, and energy transfer  \citep{Zweibel2009, Yamada2010, Pontin2011, Shibata2011, Daughton2012}. If energy is to be released through the process of magnetic reconnection, then it has to be stored in the magnetic field in the initial stage, so reconnection must be an off/on process, and cannot be occurring on any kind of \textit{fast} timescale all of the time. If this were not the case, stars and accretion disks would not have coronae, the magnetic dynamo would not work, and there would be no supersonic solar wind e.g. \citep{Zweibel2009, Yamada2010}. A complete understanding of magnetic reconnection therefore requires explaining how energy accumulates in the magnetic field, how current carrying fields becomes unstable, and how magnetic energy release occurs on short timescales once the reconnection process has been triggered.

A major difficulty in understanding of the magnetic reconnection stems from the fact that classical models of reconnection starting from the steady state Sweet-Parker mechanism \citep{parker1957, sweet1958}, or the non-steady, resistive instabilities \citep{Furth1963} appeared to be inadequate to explain the observed, transient and explosive release of the magnetic energy in various plasma environments. The fast steady state mechanism proposed by Petscheck, with a short diffusive region emanating slow mode shocks was shown in numerical simulations to depend intrinsically on non-uniform or anomalous local resistivities and impossible to achieve with quasi-uniform plasma parameters. Because reconnection is a locally small scale phenomenon strongly influenced by global conditions, it remains a difficult topic to fully understand. The extremely large values of the magnetic Reynolds and Lundquist numbers of high-temperature plasmas mean that the scales where reconnection occur may become so small that the mechanism allowing magnetic field topology change may not be tied to resistivity at all but to kinetic effects \citep[e.g.][]{Singh2015, DelSarto2016, Pucci2017}.

As had been pointed out already by \citet{Biskamp1986}, the Sweet-Parker (henceforth SP) current sheet becomes unstable in numerical simulations with sufficiently high resolution, to a very fast reconnecting mode. Here higher resolution corresponds to higher Lundquist numbers and when these Lundquist numbers became sufficiently high, the effect of inflow/outflow on the current sheet becomes negligible compared to the effect of the tearing mode \citep{Shi18}.This instability of the SP sheet \citep{Tajima2002, Loureiro2007} was called the super-tearing mode and then the plasmoid instability, as it leads both to a growth rate that \emph{increases} with the Lundquist number, and to a large number of magnetic islands, or plasmoids, being formed. Many further works were devoted to studying the SP plasmoid instability \citep[e.g.][]{huang_2010, huang_2011, Loureiro2013, huang_2013}, investigating its formation and the scaling of the number of islands formed with Lundquist number.

\citet{PV14}, hereafter PV14, noted that the main result for tearing on SP sheets, namely that the growth rate increased with increasing Lundquist number, would lead to a catastrophe in the ideal limit. PV14 therefore conjectured that in the \textcolor{red}{`}ideal' limit of high Lundquist numbers, as a current sheet thinned, magnetic reconnection would survive, but the tearing mode growth rate would become at most independent of Lundquist number. In other words, current sheets as thin as SP would never form, but reconnection would occur at thicker aspect ratios. PV14 also discussed their neglect of flow structure in the stability analysis: in their reasoning, only current sheets with SP aspect ratios \emph{require} consideration of flows, as they are required to sustain the sheets that would otherwise \emph{diffuse} away on an ideal timescale.

Defining the Lundquist number $S = LV_{\rm a}/\eta$, with $L $ the sheet half-length, $V_{\rm a}$ the Alfv\'en speed and $\eta$ the magnetic diffusivity, a SP sheet has an inverse aspect ratio $a/L\sim S^{-1/2}$, while \citetalias{PV14} found that tearing becomes 'ideal' at $a/L\sim S^{-1/3}$. In addition, they pointed out that the non-linear dependence of the growth rate on current-sheet aspect ratio could explain several phenomena in which magnetic reconnection exhibits an explosive character, in the sense that magnetic energy can be stored over a long period of time and then suddenly released on a time scale comparable with the \textit{macroscopic} ideal Alfv\'en time \citep{Tenerani2016}. In fact, \citet{Tenerani2015a} extended the analysis of \citetalias{PV14} and studied the role of viscosity on the tearing mode instability of  thin current sheets. The scalings found by \citetalias{PV14} are modified in the presence of viscosity, that allows thinner sheets to remain stable.

\citet{Shibata2001} developed the plasmoid-induced reconnection model, considered fractal tearing and found a similar criterion $a/L \le S^{-1/3}$ for the fast tearing instability following a different line of thought. Inspired by the fact that the stationary SP current sheet is stable at small Lundquist number precisely due to the outflow of material accelerated to the Alfv\'en speed along the sheet \citep{Shi18}, the question that \citet{Shibata2001} asked is for what aspect ratio the growth time of the instability timescale $\tau_g$ becomes shorter than the evacuation time along the sheet $\tau_g\le L/V_{\rm a}$.  Because the evacuation time does not depend on the Lundquist number, equality is obtained when the growth rate (or time) also becomes independent of the Lundquist number, and therefore the limiting criterion yields the same aspect ratio scaling as \citetalias{PV14}.

\citet{Shibata2001} and \citet{Singh2015} pointed out that the magnetic reconnection is  strongly time dependent and bursty, and the role of fractal-like tearing is to produce a very thin current sheet with a microscopic scale of the order of the  ion- Larmor radius or the ion inertial length. The main energy release, however, is explained by a recursive fast reconnection process which occurs after the ejection of the large scale plasmoid: this fractal magnetic reconnection model of \citet{Shibata2001} suggests that the impulsive bursty regime of reconnection is  associated with a series of plasmoids formation and subsequent ejections on various scales; in fact, \citet{Nishizuka2010} report seven plasmoid ejections associated with an impulsive burst of Hard X-Ray emission. The time-dependent nature of magnetic reconnection was noticed in three-dimensional in MHD  simulations by \citet{Nishida2013}: it was found that a Sweet--Parker type steady current sheet (at low Lundquist number) is formed below a rising flux rope. The thinning of the current sheet continues due to the rising of the flux rope until the current sheet becomes sufficiently thin so that it becomes either unstable for the tearing instability or the anomalous resistivity  sets in. During this time, the current sheet is fragmented into several small-scale current sheets, with multiple x-lines and o-lines, where current density is present as well as locally enhanced.

\citet{Tenerani2015b}, in simulations of a collapsing sheet aimed at testing the \citetalias{PV14} critical aspect ratio also observed nonlinear recursive evolution of collapsing x-points and, inspired by the \citet{Shibata2001} model, developed a similar but different analytical description of the recursive collapse. A recent review on the instability of current sheets and triggering of fast magnetic reconnection can be found in \citet{Tenerani2016}.

More recently, it has been shown how the critical aspect ratio scalings of \citetalias{PV14} change when equilibrium configurations different from the Harris current sheet profiles are considered  \citep{Pucci2018}. In the present work, we first briefly summarize the properties of the tearing instability of thin current sheets for arbitrary aspect ratios in the second section. In the third section,  we  compare the fractal reconnection scenarios developed by \citet{Shibata2001} and \citet{Tenerani2015b}, which are then generalized in section three by incorporating arbitrary current sheet profiles in the stability calculations and recursive relations. Finally, the conclusions discuss the possible implications for models of turbulent reconnection as well as three-dimensional effects.

\section{A summary of the tearing mode for current sheets with general gradients}
The dispersion relation for the reconnecting instability of a one-dimensional current sheet structure in which the magnetic field reverses sign, i.e. is odd across the sheet, while the current is even, depends, in resistive magnetohydrodynamics (MHD), on the magnetic diffusivity $\eta$, the shear-scale $a$ defining the current sheet thickness, the wavenumber $ka$ and the equilibrium structure, aligned here with the $x$-direction i.e. the $\hat{i}$-direction  and dependent only on the $y$-coordinate i.e. the $\hat{j}$- direction, defined through the relation
\begin{equation}
\label{harris}
\vec B(y) = B(y)\hat{i}= B_0 F \left( \dfrac{y}{a}\right)\hat{i},
\end{equation}
where $F$ is an arbitrary odd non-dimensional function whose first derivative provides the current profile.  $B_{0}$ is an estimate of the maximum field strength (for the Harris current sheet $ F={\tanh}(y/a)$, $B_{0}$ is also the value of the field far from the sheet). The linear stability (for incompressible fluctuations) does not depend on the presence or absence of a magnetic field in the third orthogonal direction ($z$ or $\hat{k}$) and whether the equilibrium is force-free or pressure balanced. The detailed profile of $F$ enters the dispersion relation by determining the famed $\Delta'$ parameter, which is the jump in the gradient of the reconnecting perturbed magnetic field component $\tilde b_y$ across the current sheet, as obtained by solving the corresponding component of the perturbed momentum equation assuming ideal MHD and a vanishing growth rate.  At large Lundquist number ($\eta\rightarrow 0$), analysis of the solutions to the linearized equations, subject to the boundary conditions that the velocity and magnetic field perturbations vanish far from the sheet, show that two regions define the solution structure: a boundary layer of thickness 2$\delta$ around the center ($y = 0$) of the current sheet, and outer regions where diffusivity and growth rate may be neglected - as stated for the $\Delta'$ calculation. For the Harris sheet case, $\Delta'$ is given by the expression
\begin{equation}
\Delta' = a \frac{b'_y(0^+) - b'_y(0^-)}{b_y(0)} =  2\left(\frac{1}{ka} - ka\right).
\end{equation}
In order to have instability, $\Delta'$ must be greater than 0. Two asymptotic expressions summarize the dispersion relation, depending on whether $\Delta' \delta/a <<1$ (small Delta prime or $\Delta'$, subscript SD)
\begin{equation}\label{eq:SD}
 \gamma_{_{SD}}\bar{\tau}_A\simeq A^{\frac{4}{5}}\bar{k}^{\frac{2}{5}}(\Delta')^{\frac{4}{5}}\bar{S}^{-\frac{3}{5}}\qquad \delta_{_{SD}}\sim
(\bar{S}\bar{k})^{-\frac{2}{5}}(\Delta')^{\frac{1}{5}},
\end{equation}
where $A$ is a non-dimensional constant, or
$\Delta' \delta/a >>1$ (large Delta prime or $\Delta'$, subscript LD)
\begin{equation}
\label{eq:LD}
\gamma_{_{LD}}\bar{\tau}_A\simeq \bar{k}^{\frac{2}{3}}\bar{S}^{-\frac{1}{3}}\qquad \delta_{_{LD}}\sim (\bar{S}\bar{k})^{-\frac{1}{3}},
\end{equation}
in which case the growth rate no longer depends explicitly on $\Delta'$. Here barred quantities are normalized to the current sheet \emph{thickness} ($\bar \tau_A = a/V_{\rm a}$, $\bar k = ka$, $\bar S = aV_{\rm a}/\eta$). The expressions above may be used to find the scaling of the fastest growing mode by assuming that both relations remain valid at the wave-number of maximum growth $k_m(\bar S)$ for sufficiently large $\bar S$. As the Lundquist number grows, the wave-number of maximum growth continues to decrease, and in the expression for $\Delta'$ the only part of interest is the dependence on wave-number as $ka\rightarrow 0$.  For the Harris current sheet this implies $\Delta' \sim 2/ka$, leading to
\begin{equation}
\label{maxtear}
\gamma \bar \tau_{\rm A} \sim \bar S^{ -\frac{1}{2}}\,,\,\,\,\frac{\delta}{a} \sim \bar S^{ -\frac{1}{4}}\,,\,\,\,
k_m a\sim \bar S^{-\frac{1}{4}}.
\end{equation}
\citetalias{PV14} rescaled the dispersion relation to current sheet half-length rather than half-thickness and times
normalized to the Alfv\'en time along the sheet (i.e., unbarring the Lundquist number and Alfv\'en time):
\begin{equation}
\label{harrisdisp}
\gamma \tau_{\rm A} \sim S^{-\frac{1}{2}}\left({\frac{a}{L}}\right)^{-\frac{3}{2}}\,,\,\,\
\frac{\delta}{L} \sim S^{-\frac{1}{4}}\left({\frac{a}{L}}\right)^{\frac{3}{4}}\,,\,\,\
k_m L\sim S^{-\frac{1}{4}}\left({\frac{a}{L}}\right)^{-\frac{5}{4}}.
\end{equation}
\citetalias{PV14} then argued, assuming an inverse aspect ratio of the form $a/L\sim S^{-\alpha}$, that any value of $\alpha<1/3$ would lead to a divergence of growth rates in the ideal limit, while any value of $\alpha>1/3$ would lead to growth rates which tend to zero as the Lundquist number grows without bound.
In order to preserve a physically consistent ideal MHD limit, \citetalias{PV14} therefore argued that $\alpha=1/3$ was a critical exponent at which
inverse aspect ratio current sheets would continue to tear, at the ideal rate, when $S\rightarrow \infty$. They also pointed out how
the nonlinear dependence of the growth rate on the aspect ratio could allow the tearing mode to provide a \emph{trigger} for non-ideal explosive events.

The expressions in eqs.(\ref{maxtear}, \ref{harrisdisp}) may be generalized to other equilibria by allowing for a different function $\Delta'$, and more specifically with its functional dependence on $ka$, in the limit of small $ka$ (the small $\Delta'$ regime).  For small values of $ka$ what is important is that we may assume $\Delta' \sim (ka)^{-p}$ with $p>0$, see \citep{Pucci2018} for an example of a current sheet with vanishing far field that leads to $p=2$, as well as a more general discussion, including for example a double current sheet and different boundary conditions). This leads to the generalized fastest growing mode dependencies
\begin{equation}
\gamma \bar \tau_{\rm A} \sim \bar S^{ -\frac{1+p}{1+3p}}\,,\,\,\,\frac{\delta}{a} \sim \bar S^{ -\frac{p}{1+3p}}\,,\,\,\,
k_m a\sim \bar S^{-\frac{1}{1+3p}}.
\end{equation}
Notice that such scalings imply that the exponents of power-law dependencies on growth rate etc. do not have a large domain of variation: the maximum growth rate scales with Lundquist number with an exponent between $-1/2$ (i.e. for $p=1$) and $-1/3$ (i.e. for $p\rightarrow \infty$) , while maximum wave-number and singular layer thickness exponents both vary between $-1/4$ (i.e. for $p=1$) and $-1/3$ (i.e. for $p\rightarrow \infty$).

Again rescaling the dispersion relation to current sheet half-length rather than thickness and normalized to the Alfv\'en time along the sheet (i.e., unbarring the Lundquist number and Alfv\'en time):
\begin{equation}
\label{genidisp}
\gamma \tau_{\rm A} \sim S^{-\frac{1+p}{1+3p}}\left({\frac{a}{L}}\right)^{-2\frac{1+2p}{1+3p}}\,,\,\,\
\frac{\delta}{L} \sim S^{-\frac{p}{1+3p}}\left({\frac{a}{L}}\right)^{\frac{1+2p}{1+3p}}\,,\,\,\
k_m L\sim S^{-\frac{1}{1+3p}}\left({\frac{a}{L}}\right)^{-\frac{2+3p}{1+3p}}.
\end{equation}
From these relations one can then find the critical exponents for the current sheet aspect ratio at which the growth rates no longer depend on the Lundquist number, i.e. for what relationship between $a/L$ and $S$ the growth rate is independent of $S$:
\begin{equation}
\label{genidtear}
\frac{a}{L} \sim S^{-\frac{1+p}{2(1+2p)}} \,\,\,,
\end{equation}
where again the exponent can only vary between $-1/3$ (i.e. for $p=1$) and $-1/4$ (i.e. for $p \rightarrow \infty$). Similarly one finds
the scalings of the maximum growth wavenumber and of the singular layer thickness with the Lundquist number $S$:
\begin{equation}
\label{deltauni}
k_m L \sim S^{\frac{p}{2(1+2p)}} \,\,\,, \frac{\delta}{L} \sim S^{-\frac{1}{2}}.
\end{equation}
From this we can remark that, while for the wavenumber of maximum growth the exponential dependence on $S$ varies between $1/6$ (i.e. for $p=1$) and $1/4$ (i.e. for $p \rightarrow \infty$),  the singular layer thickness (or inverse aspect ratio) is universal and equilibrium independent, scaling precisely as the SP current sheet inverse aspect ratio. The fact that the singular layer thickness is universal and equilibrium independent may seem surprising at first sight, but is actually a simple consequence of the requirement that the tearing mode instability proceed on ideal timescales. As remarked by \citetalias{PV14}, such an aspect ratio is the only one that allows dissipation to balance the perturbed inflowing magnetic energy, even in the presence of a growing mode. It might also be one of the reasons numerical simulations tend to identify sheets with inverse aspect ratios scaling as SP, though the ratio of outflow to inflow velocities does not satisfy the proper SP scaling. Indeed it is easy to see that incompressibility implies that the ratio of outflow to inflow velocities at the x-point scales, in the presence of an ideally growing tearing mode, as
\begin{equation}
\label{vinout}
V_{\rm in}/V_{\rm out} \sim S^{-\frac{(1+p)}{2(1+2p)}}
\end{equation}
and is larger than the stationary SP inflow/outflow ratio, scaling with an exponent monotonically decreasing with $p$ between $1/3$ and $1/4$. In other words, for a growing tearing mode there is a \emph{larger} ratio of inflow to outflow, as required by an exponentially increasing reconnection rate in the linear phase of the instability.

\section{Fractal reconnection scenarios compared}

 The scenario where a fast reconnection instability would disrupt current sheets whose inverse aspect ratio was small enough had already been observed in numerical simulations by \citet{Biskamp1986}. This process was discussed in more in detail in terms of multiple plasmoid formation by \citet{Tanuma1999, Tanuma2001} and \citet{Shibata2001} who used the term \textit{secondary tearing instability}.  For Sweet--Parker current sheets \citep[e.g.][]{Samtaney2009, Loureiro2013} provided studies showing their violent instability to super-Alfv\'enic fast formation of plasmoid chains. Indeed it was the singularity and non-causality implied by the Lundquist number scaling that lead \cite{PV14} to conclude that the Sweet--Parker current sheet may not form in the beginning \citep{Lapenta2008, Shibata2016}.  The presence of many plasmoids in a long and thin current sheet makes the current sheet more unstable and the overall  reconnection process evolves in a strongly time-dependent manner \citep{Tajima2002}. The recursive reconnection scenarios that we compare here are all based upon the realization that Sweet-Parker sheets are never attained, but rather a scenario we have called Ideal Tearing (IT) holds \citep[e.g.][]{ Tenerani2015b, Tenerani2016}.

\citet{Tanuma1999, Tanuma2001, Tanuma2005} conducted several MHD based numerical experiments on reconnection
in current sheets, including different models for the resistivity; the fundamental results from these simulations are summarized below:\\
\\
(i) The initial current sheet considered in the beginning of the simulation (or reconnection) is macroscopic, not a stationary SP current sheet,  and the current sheet evolves by thinning from the start of the simulation. \\
(ii) When magnetic reconnection begins, in the presence of uniform resistivity, reconnection appears to be well described by the Sweet--Parker scaling.\\
(iii) Once the thinned current sheet becomes unstable to tearing, plasmoid formation occurs and smaller scales are achieved. If anomalous resistivity is allowed to set in (i.e. local regions of lower Lundquist number), reconnection becomes Petschek like.\\

The numerical results of \citet{Tanuma1999, Tanuma2001}, inspired the fractal reconnection model described by \citet{Shibata2001}: starting from the observation that magnetic reconnection occurs in a strongly time dependent and nonlinear manner, it describes an overall process that proceeds through stages during which the current sheet aspect ratio keeps changing before successive plasmoid formation in the current sheet, at smaller and smaller scales, set in.

\subsection{The Shibata Fractal Reconnection model}
The \citet{Shibata2001} model starts from Harris-type current sheet profiles ($p=1$ in the notation of the previous section), observing that for a current sheet in which inflows and outflows are present, instability requires that the tearing mode timescale (corresponding to maximum growth rate) should be smaller than the time required for the outflow to carry the perturbation out of the current sheet. Tearing of the current sheet occurs once the instability criterion is satisfied (see eq. 12). Using the same notation as in the previous section, barred non-dimensional quantities are normalized using the current sheet half-thickness $a$, while non barred quantities are normalized to the current sheet half-length $L$. Consider a current sheet that breaks up, with x-points collapsing to give rise to secondary current sheets, in a step- by- step process until kinetic or dissipation scales are reached. The thicknesses, Lundquist numbers, Alfv\'en times, etc. at the $n^{\rm th}$ step of this cascade will be denoted with the subscript $n$.

In the \citet{Shibata2001} scenario, the current sheet at the $n^{\rm th}$ step becomes unstable to plasmoid instability if
\begin{equation}
\label{flowineq}
t_n \leq L_{n} / V_{\rm a}~,
\end{equation}
where $t_n=1/\gamma_n$ is the maximum growth time of the tearing mode instability; from eq.(\ref{maxtear})
\begin{equation}
\gamma_n \sim \bar S_n^{-1/2} \bar \tau_{An},
\end{equation}
upon renormalization, as in the previous section, the inequality eq.(\ref{flowineq}) leads to
\begin{equation}\label{thickness}
\frac{a_n}{L_n} \leq S_n^{\rm -1/3},
\end{equation}
note the similarity to \citetalias{PV14} and eq.(\ref{genidtear}) with $p=1$ for the upper bound of the inverse aspect ratio:  the Sweet-Parker current sheet satisfies the inequality but with a much smaller inverse aspect ratio. Derived by comparing the Alfv\'en time (i.e. $L_n/V_a$) with tearing time (corresponding to maximum growth rate of tearing mode instability), at a particular step \citep[c.f.][]{Shibata2001}, the upper limit corresponds to a process occurring on the ideal timescale of current sheet evacuation, on the assumption that an Alfv{\'e}nic outflow is present.

At the next step, in the \citep{Shibata2001} model the length of current sheets newly born out of the x-points is fixed by the most unstable wavelength of the tearing mode in the previous step. Consider, if eq.(\ref{thickness}) is satisfied, the expression for the fastest growing wave-number i.e. eq.(\ref{deltauni}) and ($p=1$): the wavelength of the fastest growing mode is given by $\lambda=2 \pi/k_m$ leading to
\begin{equation}
L_{n+1} = {\pi} L_n S_n^{-1/6}.
\end{equation}
The number of x-points (and o-points, each of which gives rise to a plasmoid) created in the first step in a critical current sheet of Lundquist number $S$ is then $n_1\le L_1/L_2= S^{1/6}/\pi$.

On the other hand, the destabilization of the new sheet occurs at a thickness such that
\begin{equation}
\frac{a_{n+1}}{L_{ n+1}} \leq S_{n+1}^{-1/3} = S_n^{-1/3} \left( \frac{L_{n+1}}{L_n} \right)^{-1/3} = {\pi}^{-1/3}S_n^{-5/18}
\end{equation}
Assuming equality we then get a recursive relation for the Lundquist number and current sheet thicknesses
\begin{equation}
\label{scales}
S_{n+1} = {\pi} S_n^{5/6}\,\, {\rm or} \,\, S_{n+1} = \left({\pi}\right)^{6(1-(5/6)^n)} S_1^{(5/6)^n},\,\,\, {\rm and} \,\, \frac{a_{n}}{a_{n+1}} = \left(\frac{1}{\pi}\right)^{(2/3)} S_{n}^{1/9}. \end{equation}

For $S_{n} = 10^{12}$, $L_{n}/L_{n+1} \simeq 32$ and $S_{n} = 10^{6}$, $L_{n}/L_{n+1} \simeq 3.2$. The ratios of current sheet width at two consecutive levels of tearing i.e. $a_{n}/a_{n+1}$ for $S_{n} = 10^{12}$ and $S_{n} = 10^{6}$ are 10 and 2.16 respectively.
This fractal process should continue until the current sheet thickness - or rather the singular layer of a tearing mode step - either reaches the first microscopic scale,  such as the ion Larmor radius or ion inertial length, or a value for the collisional Lundquist number sufficiently small so as to stabilize tearing completely, i.e. a value $S_N\simeq 10^4$ or less \citep{Shi18}. The eq.(\ref{scales}) shows that starting from a Lundquist number
$S_1=10^{12}$ it takes 13 (12 excluding the initial sheet) steps to get to  $S_{13}\simeq10^4$.

The expected total number of x-points generated after $M$ steps then becomes
\begin{equation}
N_{M} =  N_{1}\cdot N_{2} \cdot N_{3} ..... \cdot N_{M} \le \frac{1}{\pi}S_{1}^{1/6} \frac{1}{\pi}S_{1}^{1/6} ....\frac{1}{\pi}S_{M}^{1/6}  =
\left(\frac{1}{\pi}\right)^{6(1-(5/6)^{M})} S_1^{(1-(5/6)^{M})}~.
\end{equation}
Calculating the total number of plasmoids requires an assumption on the relationship between x-points at each step, o-points, and the behaviour of islands generated at previous steps. In principle, one should sum over $N_M$ from $M=1$ through a final step to obtain an upper limit on the number of plasmoids. Here we will consider $N_M$ to be a reasonable estimate for the total number of islands formed, given that larger islands will merge during the cascade.
Assuming the cascade reaches down to $M=13$ before it stabilizes we get a total number of plasmoids $N_{13} \simeq 1.5 \times 10^8$, an enormous number. Using slightly smaller values of the initial Lundquist number led \citet{Shibata2001} to conclude that the number of steps should be $M\ge6$ (or 5, starting from an initial sheet with $M=1$). The number of plasmoids as a function of the number of steps and $S_{1}=10^{11}$, is shown in Figure 2.

\subsection{The Tenerani et al. Fractal Reconnection Model}

\citet{Tenerani2015b,Tenerani2016} carried out a series of simulations of collapsing current sheets of the Harris type with the aim of testing the \citetalias{PV14} critical aspect ratio criterion. To this end, \citet{Tenerani2015b,Tenerani2016} introduced a slowly collapsing primary sheet, and indeed observed that, once the aspect ratio thinned to a scale of order $a/L\sim S^{-1/3}$ a fast (plasmoid) type instability occurred. They observed the subsequent collapse of secondary x-points into further current sheets, and, inspired by \citet{Shibata2001} went on to model the recursive collapse. The main difference however was that \citet{Tenerani2015b} observed that collapsing x-points had the property of maintaining their \emph{thickness} unaltered while the outflow tended to lengthen them, until their inverse aspect ratio became sufficiently small that a new tearing instability was triggered. An example from their MHD simulations is shown in Figure 1, where magnetic field lines and current density are shown at $t=26 \tau_A$, $t=27 \tau_A$, $t=27.37 \tau_A$ in the evolution. At $t = 26 \tau_A$, there are many plasmoids in a long and thin current sheet. As time goes on, the current sheet lengthening is observed at $t = 27 \tau_A$, e.g. for $x/L$ between 2.7 and 3.7. At $t = 27.37 \tau_A$, there are plasmoids merging can be seen clearly, e.g. for $x/L$ between 1.7 and 2.5. So, in proceeding from the top to bottom panels, we see x-point collapsing, current sheets lengthening, and subsequent collapse, while the plasmoids formed at previous steps grow and merge.

\begin{figure}
\label{plasm1}
\centering{
\includegraphics[scale=0.6]{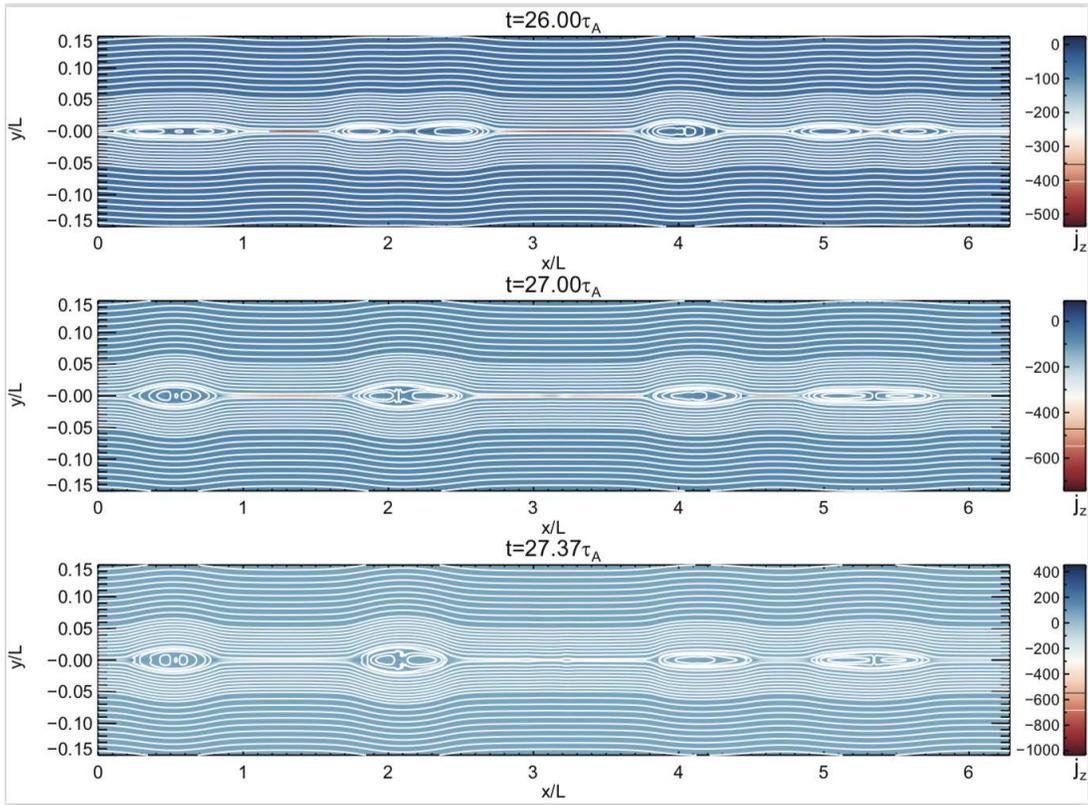}
\caption{Magnetic field lines and current density are shown at three moments i.e. for $t = 26 \tau_A$, $t = 27 \tau_A$ and $t = 27 \tau_A$ in a collapsing current sheet, from \citet{Tenerani2015b}. In the top panel, the central current sheet is lengthening (see e.g. for $x/L$ between 2.7 and 3.7). In the middle panel, the current sheet lengthening has given rise to the plasmoids. In the bottom panel, the x-points between the plasmoids have begun to collapse into a thinner current sheet.  The times are normalized to the Alfv{\'e}n time along the sheet, x, whose total length is 2$\pi$L here.}}
\end{figure}

Each subsequent x-point collapse was identified as the next step in the Ideal Tearing cascade because the new aspect ratios empirically satisfied the IT criterion with Lundquist numbers calculated on the new current sheet lengths.
In other words, the thickness of the current sheet developing at the $n+1$-th stage of the instability, corresponded to the inner, singular, diffusion layer of the $n$-th tearing mode, i.e. $\delta_{n}$ to be obtained from eq.(\ref{deltauni}) with $p=1$: as new x-points collapse, they form sheets that broaden essentially at the \emph{upstream magnetic field Alfv\'en speed}, and the current sheet thickness at stage $n+1$ is given by $a_{n+1} =\delta_n$. In order to become ideally unstable, the current sheets have to  reach a length $L_{n}$, such that their growth rate is independent of the Lundquist number calculated on the upstream magnetic field Alfv\'en speed and on the new half-length $L_n$, so, that
\begin{equation}\label{scale_ratio}
\frac{a_{n}}{L_{n}} \sim S_{n}^{-1/3},
\end{equation}
and recalling the definition of $a_n$, one finds
$a_{n+1} / L_{n+1} = \delta_{n} / L_{n+1} = (L_{n}/L_{n+1}) \cdot (S_{n}^{-1/2}) = S_{n+1}^{-1/3}$.
Note that because the IT tearing is a scaling relation, equality is up to a constant $c$ of $O(1)$ appearing in front of the expressions
(i.e. we should write $a_{n} / L_{n} = c_{n} S_{n}^{-1/3}$ with $c_n \sim O(1)$). We omit such constants here in favor of
a less cumbersome notation.

From the definition of $S_n$ and assuming the diffusivity and the upstream magnetic field to remain the same at each level (as we have done implicitly in the previous section as well), we have (our initial sheet is labelled by $n=1$)
$S_{n+1} = (L_{n+1}/L_{n})\cdot S_{n}$, so that
\begin{equation}\label{new_S}
(L_{n}/L_{n+1}) \sim S_{n+1}^{1/3} \rightarrow S_{n} = S_1^{(3/4)^{n-1}},\,\,\, n\ge1.
\end{equation}
The present model, inspired by the observed behaviour of collapsing x-points in numerical simulations, leads to interesting differences from the original \citep{Shibata2001} ideas: firstly, the Lundquist number decreases at a much faster rate in each subsequent step, because the geometric progression in the exponential has a ratio $3/4$ rather than $5/6$. In order for the Lundquist number to fall from say, $S=10^{12}$ to below $10^4$, only 5 steps are required (4 from the initial unstable sheet), i.e. $S_5=S_1^{(3/4)^4} \simeq 6264$, rather than the previous 9. The total number of plasmoids is also much reduced. In the simulations of \citet{Tenerani2015b, Tenerani2016}, it was seen that the collapse of x-points does not occur uniformly along the sheet, rather, there is a competition between plasmoid coalescence on the one hand and x-point collapse, in correspondence of the strongest currents, on the other. While for the Shibata model it was obvious that there was originally enough space along the sheet for \emph{all} of the x-points to collapse and thin to a secondary instability by construction, as the sub-current sheets were of a defined length that summed to the original one, this is not obvious for the Tenerani model. Here we ask that the total number of plasmoids in the half-length, ($N/2$), times the new instability half-length $L_{2}$ must fit inside the original sheet half-length $L_1$. Combining Eq. \ref{deltauni} and Eq. \ref{new_S},
we obtain
\begin{equation}\label{N_relation}
 N/2 \cdot L_{2} \simeq S_1^{1/6}/ (2 \pi) \cdot L_1 S_1^{-1/4} \leq L_1,
\end{equation}
 \begin{center}
 $ 1/(2 \pi) S_1^{\rm -1/12} \leq 1$,
 \end{center}
 which is always satisfied for $S_1>1$. The maximum total number of x-points (and therefore plasmoids) that could form in this fractal collapse is given as before by the product of those produced at each step: the total number after $M$ ($M$-1 secondary)-steps becomes
\begin{equation}
\label{count1}
N_{M} = N_{1} \cdot N_{2} \cdot ..... \cdot N_{M} = \left({1\over \pi}\right)^{M}  S_{1}^{1/6} S_{2}^{1/6} ....S_{M}^{1/6} =
\left({1\over \pi}\right)^{M}S_1^{2/3(1-(3/4)^{M})}~.
\end{equation}

Again, from a realistic viewpoint, $M$ is limited by stabilization at small $S$, whether due to velocity field or otherwise, somewhere around $S_{M} \simeq 10^{\rm 4}$, so starting from $S_1 = 10^{\rm 12}$, we find that $M$ is at most 5 (actually, 4.8); this leads to a number of islands
$N_{M} \geq 0.003  \cdot S_1^{0.5}$. However, if we assume that only some fraction i.e. $f_i$ of number of plasmoids always make themselves available for the tearing mode instability at the next step, then this number is reduced by a factor $\Pi_{i=2}^M f_i$. Although $f_i \le 1$, but for the contribution to the number of plasmoids, here $f_i$ is limited by the fact that $f_i n_{i+1} \ge 1$, and for a given value of $S = 10^{\rm 12}$, $n_1 \simeq 32$, $n_2 \simeq 10$, $n_3 \simeq 4$, $n_4 \simeq 2$, $n_5\simeq1$.
This means that starting from $S = 10^{\rm 12}$, $M=5, f = 0.1$, a reasonable estimate of the total number of plasmoids is $N_{M} \geq 3$. The number of plasmoids as a function of number of steps, and based upon \citet{Tenerani2015b} and $S_{1}=10^{12}$, is shown in Figure 2.

\citet{Tenerani2016} showed that in the hierarchical collapse of current sheets the upstream magnetic field did not tend to be reduced
significantly during each step in the self-similar collapse. However the possibility of a reduced upstream field should not be discounted,
so we consider its effect here.  In this case, the reconnecting field is not the upstream field $B_{\rm 0}$ (defining $V_{\rm a}$ in the Lundquist number) but only some fraction $\beta$ of that value. Changing $V_{\rm a}$ to $\beta V_{\rm a}$ and consequently renormalizing all quantities,  the main difference in the number of islands comes via this Lundquist number reduction, each one of which gaining a factor $\beta$, leading to a modification of the scaling relation eq.(\ref{new_S}) into
\begin{equation}
S_n= \beta^{6(1-(3/4)^{n-1})}S_1^{(3/4)^{n-1}}
\end{equation}
and a generalization of eq.(\ref{count1}) yielding the number of x-points or islands after $M$ -steps into
\begin{equation}
N_{M} = \left({1\over \pi}\right)^{M}S_{1}^{1/6} S_{2}^{1/6} ....S_{M}^{1/6} = \left({\beta\over \pi}\right)^M \beta^{-4 (1-(3/4)^{M} ) }S_1^{2/3(1-(3/4)^{M})}~.
\end{equation}

\subsection{Generalized Scaling laws for Recursive reconnection (Ideal tearing)}

Although \citetalias{PV14} gave a compelling argument against the requirement of including flows in their study of the tearing in isolated sheet, during the recursive tearing scenario flows are indeed present, and such flows, following the arguments of \citet{Shibata2001}, might lead to a slightly different exponent $\alpha\le1/3$ for the critical inverse current sheet aspect ratios, even considering the $p=1$ Harris profile.  Generalizing the \citet{Tenerani2015b} model to arbitrary $\alpha$, we find, from eq.(\ref{scale_ratio}),

\begin{center}
$a_{2} / L_{2} = \delta_1 / L_{2} = (L_1/L_{2}) \cdot S_1^{\rm -(1+3\alpha)/4} = S_{2}^{-\alpha}$, ~ $S_{2} = (L_1/L_{2}) \cdot S_1$,
\end{center}
implying that
\begin{center}
$S_1^{(\alpha-1)/4} = (L_{2}/L_1)^{(1-\alpha)} \rightarrow (L_{2}/L_1) = S_1^{-1/4}$,
\end{center}
just as before. In other words, it remains true that
\begin{equation}
S_{n+1} = S_1^{(3/4)^n}
\end{equation}
However, the other scalings, such as the number of islands with Lundquist number, change:
using $a/L \sim S^{-\alpha}$ in eq.(\ref{harrisdisp}) leads to
$kL \sim S^{\rm (-1+5\alpha)/4}$, and, following the previous
procedure, the total number of islands scales as
\begin{equation}
N_{M} = \left({1\over  \pi}\right)^{M}S^{(5\alpha -1)(1-(3/4)^{M})}.
\end{equation}
From here and requesting that we have more than one island we see that $\alpha$ must obviously satisfies $\alpha>1/5$ (or the exponent of $S$ would be negative) so it is safe to assume that $1/5<\alpha<1/2$. The number of steps is determined only by the values of $S$ so it remains fixed at around 4-5 for any realistic case. Because this calculation only makes sense if the number of islands is $\geq 1$ to begin with, the calculation from the \citet{Pucci2018} gives $kL \sim S^{\rm (-1+5\alpha)/4}$ which for $S\simeq 10^{\rm 12}$ means that in reality $\alpha \geq 0.233$ in order for the process to begin.

The independence of the exponential geometric progression in $S$ on the coefficient $\alpha$ is one interesting aspect. The generalization
allowing for magnetic field embedding (i.e.  the factor $\beta$ in the Lundquist number at step $i>1$ leads again to
$S_{n+1} = \beta (L_{n+1}/L_{n}) S_{n}$ and $S_{\rm n} = S^{\rm (3/4)^n} \beta^{6 (1-(3/4)^n)}$.

The total time ($\tau_{M}$) for recursion to reach microscopic scales, once the first step has been triggered, i.e. from $n=2$ to $M$ is \begin{center}
$\tau_{M}\simeq \sum\limits_{n=2}^{{M}} \tau_{A,n}~,$
\end{center}
where $\tau_{A,n} = (L_{n}/L_1) \tau_{A1}$. For $\alpha=1/3$ and neglecting the dependence on $\beta$ this leads to
\begin{center}
$\tau_{M} = S_1^{-1/4} \tau_{A1} [1+S_1^{-1/4(3/4)}+S_1^{-1/4(3/4+(3/4)^2)}+.... ] \sim S_1^{-1/4} \tau_{A1}.$
\end{center}
For $S=10^{12}$ that is typical of solar corona, $\tau_{5} \sim (5 \times 10^{\rm -4}) \tau_{\rm A}$. Any other value of $\alpha<1/3$ leads to timescales which are faster.

Let us now come back to the same question, but this time instead of using an arbitrary $\alpha$ and a Harris current sheet, consider different
profiles with arbitrary $p$. The generalization of recursive tearing following \citet{Tenerani2015b, Tenerani2016} is simple, following the same steps as above but using eqs.(\ref{genidtear}, \ref{deltauni}). The recursive relation for the Lundquist number $S$ now reads
\begin{equation}
\label{genL}
S_{n+1}= S_1^{\left(\frac{1+2p}{1+3p}\right)^n}
\end{equation}
while the number of x-points generated (assuming that recursive collapse occurs at all x-points) becomes
\begin{equation}
\label{genisl}
N_{M} = \left({1\over \pi}\right)^M S_1^{\frac{1}{2}\left(\frac{1+3p}{1+2p}\right)\left(1-(\frac{1+2p}{1+3p})^M\right)}
\end{equation}
Note that changing the value of $p$ does not change the recurrence dramatically, as the exponents vary monotonically between $(3/4)^n$ (i.e. for $p=1$) and $(2/3)^n$ (i.e for $p\rightarrow \infty$) in the Lundquist number eq.(\ref{genL}) and between $2/3$ ($p=1$) and $3/4$ ($p\rightarrow \infty$) in the power in front of the parenthesis in the exponent of eq.(\ref{genisl}).

\begin{figure}
\label{plasm2}
\centering
\includegraphics[scale=0.9]{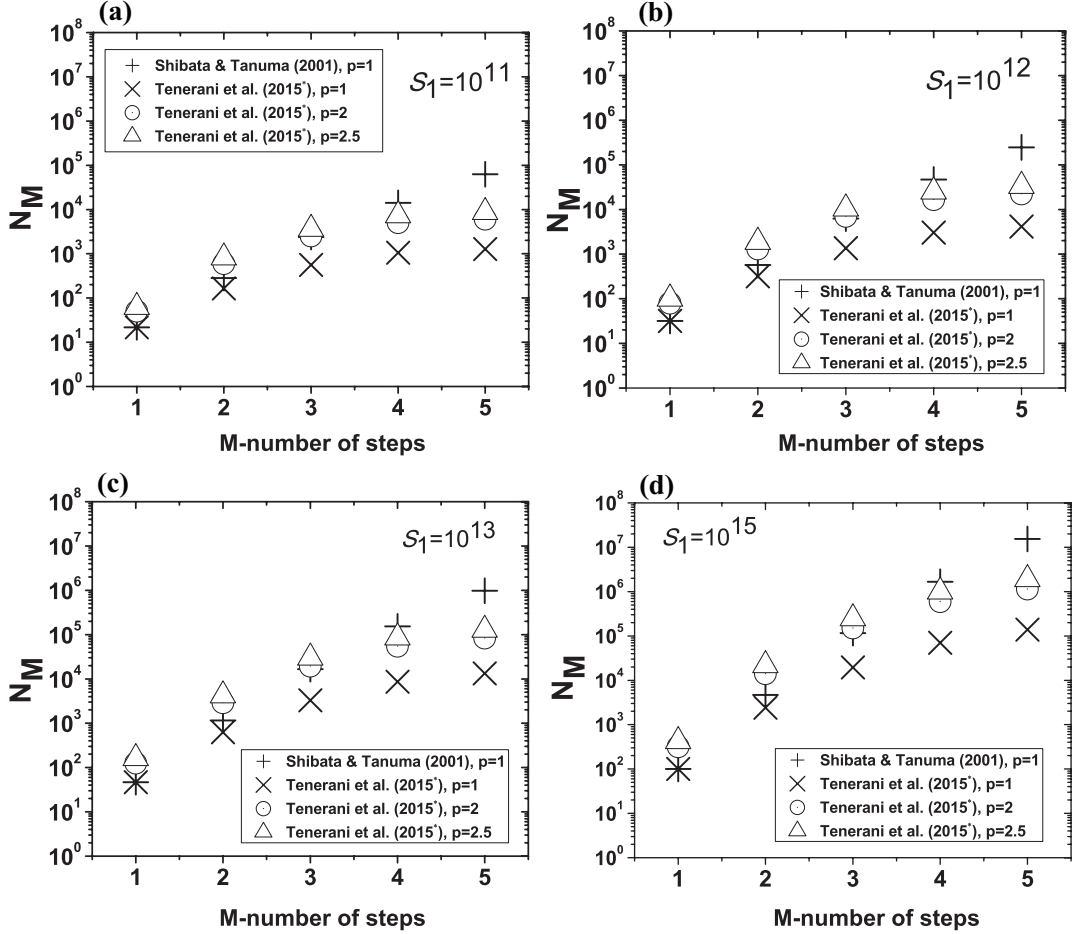}
\caption{ The number of plasmoids ($N_{\rm M}$) as a function of $M$. The calculations for the number of plasmoids are based upon \citet{Shibata2001} and \citet{Tenerani2015b}. (a) for $S_{\rm 1}= 10^{11}$, (b) for $S_{\rm 1}= 10^{12}$, (c) for $S_{\rm 1}= 10^{13}$, and (d) for $S_{\rm 1}= 10^{15}$. In the figure, $\star$-symbol in front of Tenerani et al. (2015) refers to \citet{Tenerani2015b}.}
\end{figure}

Further generalizations follow by relaxing some of the hypotheses considered up to now. Again using $p=1$, and  $\alpha$ for the scaling of the inverse aspect ratio with Lundquist $a/L \sim S^{-\alpha}$, eq.(\ref{harrisdisp}) shows that the singular layer thickness will scale with exponent
$\alpha'$, such that $\alpha'= \alpha'(\alpha) = -(1+3\alpha)/4$. In our stepwise process, we have considered the thickness $a_{n+1}$ to be the inner diffusion layer of the $n^{\rm th}$ tearing step, $\delta_{n}$
in the recursive reconnection based upon the ideal tearing scenario of \citetalias{PV14}. For a generic $\alpha$, the equations:
\begin{equation}
\frac{a_{n+1}}{L_{n+1}} \sim S_{n+1}^{-\alpha}~,
\end{equation}
and
\begin{equation}
\frac{a_{n+1}}{L_{\rm n}} \sim S_{\rm n}^{-\alpha'}~,
\end{equation}
would yield the following scaling laws:\\
\begin{equation}
\frac{L_{n}}{L_1}= S_1^{\tfrac{\alpha-\alpha'}{1-\alpha}\left[(1-\chi^n)/(1-\chi)\right]}~,
\end{equation}
\begin{equation}
S_{n} = S_1^{\rm (1+\varsigma)}.
\end{equation}
Here $\chi = (1-\alpha')/(1-\alpha)$ and $\varsigma = \frac{(\alpha-\alpha')}{(1-\alpha)}[(1-\chi^n)/(1-\chi)]$. So, the scalings derived in eqs. (31) and (32) depend on one parameter only.

Consider now a further possibility $L_{2}/L_1 \sim S_1^{-\beta'}$, expressing
the fact that in a secondary collapse the resistivity $\eta_{\rm 1} \neq \eta$ but rather a function $\eta_{\rm 1}=\eta_{\rm 1}(\eta)$.
We want to investigate what changes in the total number of islands.\\
The condition which expresses the total number of x-points ($N/2$) that should fit-in, inside the original half-length $L$ for $\alpha=1/3$ is given in Equation \ref{N_relation}.
For a generic $\alpha$ this condition becomes:
\begin{equation*}
\pi N \simeq  kL_2 \sim S_1^{-1/4} S_1^{5\alpha/4} \simeq S_1^{1/4(5\alpha-1)} \implies
\end{equation*}
\begin{equation}
\frac{S_1^{1/4(5\alpha-1)}}{2\pi}\frac{L_{\rm 2}}{L_1} \leq 1~.
\end{equation}
Now, if we generalize $L_{\rm 2}/L_1 \sim S_1^{\rm -\beta'}$ (instead of $L_{\rm 2}/L_1 = S_{\rm 2}/ S_1$) we get
\begin{equation}
S_1^{\rm (5\alpha-1-4\beta')} \leq (2\pi)^4 \simeq 10^3~,
\end{equation}
so that $\beta'=\beta'(\alpha)$. Now suppose $S=10^{\rm 12}$, then
\begin{equation}
5\alpha-1-4\beta' \leq 1/4 \implies \beta' \geq \frac{5}{4}\left(\alpha-\frac{1}{4}\right)~.
\end{equation}
Finally, we can compare the value of $\beta'$ for the case of plasmoid instability \textit{in a Sweet-Parker current sheet} ($\alpha=1/2$)
and for the IT case ($\alpha=1/3$)
\begin{equation}
\alpha=1/2 \implies \beta' \geq 5/16
\end{equation}
\begin{equation}
\alpha=1/3 \implies \beta' \geq 5/48~,
\end{equation}
so that in the SP plasmoid case $L_{\rm _2}/L_1 \leq S_1^{\rm -5/16}$ which means, since we considered $S=10^{\rm 12}, L_{\rm 2}/L_{\rm 1} \leq 10^{\rm -4}$, while in the ideal tearing plasmoid mode case $L_{\rm 2}/L_1 \leq S_1^{\rm -5/48}$, so $L_{\rm 2}/L_1 \leq 0.03$. A secondary collapse in the IT instability is numerically easier to follow in terms of
resolution with respect to the plasmoid case.

\section{Summary and Conclusion}

Plasmoid-mediated reconnection plays an important role in reconnection dynamics. \citetalias{PV14} proposed the ideal tearing model, in which the tearing mode grows with Alfv\'en timescale at a critical inverse aspect ratio (i.e. $a/L \sim S^{-1/3}$ for the Harris sheet). \citet{Tenerani2015b} developed a magnetic reconnection model describing the nonlinear recursive evolution. Here we have generalized  the scalings derived by \citetalias{PV14} and \citet{Tenerani2015b} both to arbitrary current profiles and to different possible stabilizing effects. The subtle differences between the recursive model based upon ideal tearing and the fractal reconnection are discussed. Depending  upon the situation \citep[e.g.][]{Tenerani2015b, DelZanna2016, Baty2017, Pucci2018}, the scalings for the fastest growing mode in a current sheet get modified.

 One of the important result is that the singular layer in an ideally unstable layer is found to be universal and it does not depend upon the details of the initial current profile in the current sheet. The scaling relations including the departure from the Harris-type initial current sheet are derived (c.f. section 3). We have also derived scaling relation for the number of plasmoids by generalizing recursive reconnection based upon the Ideal Tearing scenario. The number of plasmoids can be compared for $p=1$, $p=2$ and $p=2.5$ (Figure 2); the number of plasmoids for $p=2$ and $p=2.5$ show departures from the Harris-type initial current sheet. The present study shows that the departure from initial, Harris type current sheet affects the plasmoid formation. The number of plasmoids formed in a recursive, ideal tearing (Tenerani et al. 2015b) is less compared to those predicted by \citet{Shibata2001}, see e.g., $M = 5$ of Figure 2. Showing the comparison between number of plasmoids for various cases, we can conclude that the plasmoid dynamics (i.e. x-point collapse combined with the plasmoid formation, merging and intermittent ejection) plays a crucial role. Therefore, we can understand that plasmoid dynamics plays an important role in the evolution of the current sheet and fast reconnection. A few studies show that the multi-plasmoid instability observed in MHD simulations often arises due from numerical resistivity, and when numerical artifacts are minimized (by increasing simulation numerical resolution), a well-defined unique value of the critical Lundquist number beyond  which the plasmoid instability is fully developed may not exist, as it depends on the precise simulation setup including the noise or fluctuations in the initial conditions, as well as boundary conditions \citep{Shimizu2017,Shi18}. Even though plasmoid instability models have become a very popular subject for the fast reconnection process, a clear understanding  of the plasmoid-mediated processes is fundamental as well as vital. The present work, provides a guide and predictions to understand the plasmoid instability and recursive reconnection for various scenarios. In addition, the model we describe has implications for reconnection modified turbulence theories \cite{Boldyrev2017} that are beyond the scope of the present paper.

A final comment concerns the nonlinear evolution in the presence of kinetic effects, such as the Hall term in the induction equation, that lead directly to an intrinsic 3D behavior. Only few simulations have been carried out in 3D, but it may well be that in the absence of a strong guide field the simple recursive picture presented here will not hold. Although plasmoids may form, their twisting and kinking in three dimensions may lead to a less structured behaviour in the subsequent nonlinear evolution \citep[e.g.][]{Landi2008}. 

\acknowledgements
KAPS gratefully acknowledges the UGC Faculty Recharge Program of Ministry of Human Resource Development (MHRD), Govt. of India and University Grants Commission (UGC), New Delhi as well as the visiting associateship program of Inter University Centre for Astronomy \& Astrophysics (IUCAA), Pune.
AH is supported by his STFC Ernest Rutherford Fellowship grant number ST/L00397X/2 and by STFC grant ST/R000891/1. A.T. and M.V. were supported by the NSF-DOE Partnership in Basic Plasma Science and Engineering award N.1619611 and the NASA Parker Solar Probe Observatory Scientist grant NNX15AF34G.\\


\begin{thebibliography}{}
\bibitem[Baty(2017)]{Baty2017} Baty, H. 2017, \apj, 837, 74
\bibitem[Biskamp(1986)]{Biskamp1986} Biskamp, D.\ 1986, Phys. Fluids, 29, 1520 %
\bibitem[Bhattacharjee et al.(2009)]{Bhat2009} Bhattacharjee, A. et al.\ 2009, PhPl, 16, 112102 %
\bibitem[Boldyrev \& Loureiro(2017)]{Boldyrev2017} Boldyrev, S. \& Loureiro, N.~F.\ 2017, \apj, 844, 125 %
\bibitem[Daughton \& Roytershteyn(2012)]{Daughton2012} Daughton, W. \& Roytershteyn, V.\ 2012, \ssr, 172, 271 %
\bibitem[Del Sarto et al. (2016)]{DelSarto2016} Del Sarto, D., Pucci, F. Tenerani, A. \& Velli, M.\ 2016, \jgr, 121, 1857 %
\bibitem[Del Zanna et al.(2016)]{DelZanna2016} Del Zanna, L. et al.\ 2016, J. Phys. Conf. Ser., 719, 012016
\bibitem[Furth et al.(1963)]{Furth1963} Furth, H.P., Killeen, J. \& Rosenbluth, M.N.\ 1963, Phys. of Fluids, 20, 459 %
\bibitem[{ Huang and Bhattacharjee(2010)}]{huang_2010} Huang, Y.-M., Bhattacharjee, A.\ 2010, PhPl, 17, 062104 %
\bibitem[{ Huang et al.(2011)}]{huang_2011} Huang, Y.-M., Bhattacharjee, A., and Sullivan, B. P.\ 2011, PhPl, 18, 072109 %
\bibitem[{ Huang and Bhattacharjee(2013)}]{huang_2013} Huang, Y.-M., Bhattacharjee, A.\ 2013, PhPl, 20, 055702 %
\bibitem[Landi et al.(2008)]{Landi2008} Landi, S., Londrillo, P., Velli, M., \& Bettarini, L.\ 2008, Physics of Plasmas, 15, 012302 %
\bibitem[Lapenta(2008)]{Lapenta2008} Lapenta, G.\ 2008, \prl, 100, 23500 %
\bibitem[Loureiro et al.(2007)]{Loureiro2007} Loureiro, N.~F., Schekochihin, A.~A. \& Uzdensky, D.~A.\ 2007, PhPl, 14, 100703 %
\bibitem[Loureiro et al.(2013)]{Loureiro2013} Loureiro, N.~F., Schekochihin, A.~A. \& Uzdensky, D.~A.\ 2013, \pre, 87, 013102 %
\bibitem[Nishizuka et al.(2010)]{Nishizuka2010} Nishizuka, N. et al.\ 2010, \apj, 711, 1062 %
\bibitem[Nishida et al. (2013)]{Nishida2013} Nishida, K., Nishizuka, N. \& Shibata, K.\ 2013, \apj, 775, L39 %
\bibitem[Parker(1957)]{parker1957} Parker, E.N.\ 1957, \jgr, 62, 509 %
\bibitem[Petschek(1964)]{Petschek1964} Petschek, H.E.\ 1964, Physics of Solar Flares, Proc. of AAS-NASA Symposium (ed. W.N. Hess), 425 %
\bibitem[Pontin(2011)]{Pontin2011} Pontin, D.~I.\ 2011, Adv. in Space Res., 47, 1508 %
\bibitem[Priest(2014)]{Priest2014} Priest, E.\ 2014, Magnetohydrodynamics of the Sun (Cambridge University Press: UK) %
\bibitem[Pucci et al.(2017)]{Pucci2017} Pucci, F., Velli, M. \& Tenerani, A.\ 2017, \apj, 845, 25 %
\bibitem[Pucci \& Velli(2014)]{PV14} Pucci, F. \& Velli, M., 2014, \apjl, 780, L19 %
\bibitem[Pucci et al.(2018)]{Pucci2018} Pucci, F., Velli, M., Tenerani, A. \& Del Sarto, D.\ 2018, PhPl, 25, 032113 %
\bibitem[Samtaney et al.(2009)]{Samtaney2009} Samtaney, R. et al.\ 2009, \prl, 103, 105004 %
\bibitem[Shi et al.(2018)]{Shi18} Shi, C., Velli, M. and Tenerani, A.\ 2018, \apj, 859, 253 %
\bibitem[Shibata \& Tanuma(2001)]{Shibata2001} Shibata, K. \& Tanuma, S.\ 2001, EP\&S, 53, 473 %
\bibitem[Shibata \& Magara(2011)]{Shibata2011} Shibata, K. \& Magara, T.\ 2011, Living. Rev. of Sol. Phys., 8, 6 %
\bibitem[Shibata \& Takasao(2016)]{Shibata2016} Shibata K. \& Takasao S.\ 2016, Fractal Reconnection in Solar and Stellar Environments.
In: Gonzalez W., Parker E. (eds) Magnetic Reconnection. Astrophysics and Space Science Library, v427. Springer, Cham %
\bibitem[Shimizu et al.(2017)]{Shimizu2017} Shimizu, T., Kondoh, K. \& Zenitani, S.\ 2017, PhPl, 24, 112117 %
\bibitem[Singh et al.(2015)]{Singh2015} Singh K.~A.~P., Hillier, A., {Isobe}, H. and {Shibata}, K.\ 2015, \pasj, 67, 96 %
\bibitem[Sweet(1958)]{sweet1958} Sweet, P.A. \ 1958, In Electromagnetic Phenomena in Cosmical Physics (Cambridge University Press), 6, 123 %
\bibitem[Tajima \& Shibata(2002)]{Tajima2002} Tajima, T. \& Shibata, S.\ 2002, Plasma Astrophysics, Westview Press (USA) %
\bibitem[Tanuma et al.(1999)]{Tanuma1999} Tanuma, S. et al.\ 1999, \pasj, 51, 161 %
\bibitem[Tanuma et al.(2001)]{Tanuma2001} Tanuma, S. et al.\ 2001, \apj, 551, 312 %
\bibitem[Tanuma \& Shibata(2005)]{Tanuma2005} Tanuma, S. \& Shibata, K.\ 2005, \apjl, 628, L77 %
\bibitem[Tenerani et al.(2015a)]{Tenerani2015a} Tenerani, A., Rappazzo, A.~F., Velli, M. \& Pucci, F. 2015a, \apj, 801, 145 %
\bibitem[Tenerani et al.(2015b)]{Tenerani2015b} Tenerani, A., Velli, M., Rappazzo, A.~F. \& Pucci, F. 2015b, \apj, 813, L32 %
\bibitem[Tenerani et al.(2016)]{Tenerani2016} Tenerani, A., Velli, M., Pucci, F., Landi, S. \& Rappazzo, A.~F.\ 2016, Journal of Plasma Phys., 82, 5 %
\bibitem[Yamada et al.(2010)]{Yamada2010} Yamada, M., Kulsrud, R. \& Ji, H. \ 2010, Rev. Mod. Phys., 82, 603 %
\bibitem[Zweibel \& Yamada(2009)]{Zweibel2009} Zweibel, E.~G. \& Yamada, M.\ 2009, \araa, 47, 291 %
\end{thebibliography}
\end{document}